\documentstyle[prd,aps]{revtex}
\begin{document}
\draft
\title{
High Energy Hadron-Nucleus Cross Sections  and Their Extrapolation to
Cosmic Ray Energies}

\author{James S. Ball and Anthony  Pantziris }
 
\address{
University of Utah, Salt Lake City, UT ~84112
}
\date{\today}
\maketitle
\begin{abstract}
Old models of the scattering of composite systems based on the Glauber model 
of multiple diffraction  are applied to hadron-nucleus scattering. We obtain an
excellent fit with only two free parameters  to the highest energy 
hadron-nucleus data available.  Because of the quality of the fit and the simplicity 
of the model it is argued that it should continue to be reliable up to the highest 
cosmic ray energies.  Logarithmic extrapolations of $p-p$ and
$\bar{p}-p$ data are used to calculate the proton-air cross sections at very 
high energy.  Finally, it is observed that if the exponential 
behavior of the 
$\bar{p}-p$ diffraction peak continues into the few TeV energy range it 
will violate partial 
wave unitarity.  We propose a simple modification that will guarantee 
unitarity throughout the cosmic ray energy region.
\end{abstract}

\section{Introduction}
The highest energy hadronic interactions currently observable occur when a primary 
cosmic ray strikes an air nucleus.  For primary protons with 
energy of $10^{18}$
eV, which are easily observed in the Utah ``Flys's Eye'' detector, 
the $\sqrt{s}$ exceeds that of the recently cancelled SSC.  The phenomenology
of the extensive air showers created by the hadronic cascade in the atmosphere 
depends critically on the hadron-nucleus and the nucleus-nucleus cross sections
at extremely high energy.  Furthermore, extracting any information about the
basic hadronic interactions requires some model that can reliably relate the 
nuclear cross section to the 
nucleon cross section.  The purpose of this paper is
to develop such a model and check it with experimental data in the energy range
in which both the hadron-nucleus and hadron-nucleon cross sections are known.
   
At high energies hadron-nucleon scattering amplitudes become very simple.  
The amplitude is diffractive being dominated by the imaginary part with rather
 weak (logarithmic) energy dependence.  The momentum transfer dependence in the
 dominant part of the diffraction peak is a pure exponential.  This is to be 
contrasted with the situation at lower energies where resonance phenomena 
produce rapid energy dependence and complicated angular dependence.   
In this paper we have resurrected an old model of Fishbane and Trefil, and of 
Franco\cite{r1} and show that this simple application of the Glauber multiple 
diffraction model\cite{r2}  can be used to provide reliable
hadron-nucleus cross sections   from  the basic hadron-nucleon
interaction.  We further argue that since no energy dependence
is required in this model in fitting the data, i.e. the distribution of 
nucleons in the nucleus is independent of the momentum of the incident hadrons,
this method should continue to be applicable to scattering at  the
highest cosmic ray energies. One important reason that the nuclear
corrections are simple at high energy is that for large $A$ the
nucleus is mostly black and only the peripheral shell of the nucleus is sensitive to the
details of the hadron-nucleon interaction; thus the nuclear cross sections depend
only weakly on the  nucleon density distributions  and hadron-nucleon
cross sections. It should be kept in mind, however, that Glauber's
approximation is only valid for small scattering angles.   

In the next section we will review the Glauber multiple diffraction  model for 
hadron-nucleus and nucleus-nucleus scattering,
assuming a Gaussian distribution of nucleons in the nucleus. In section III,
 we determine the input parameters, the hadron-nucleon cross sections and slope
parameters for $p-p, \bar{p}-p, \pi^{+}-p, \pi^{-}-p, K^{+}-p$ and $K^{-}-p$.  The
two parameters of the nuclear model are then adjusted to fit the corresponding
 nuclear cross sections as measured by Carroll et al. \cite{r3} 
 
In section IV we obtain  simple fits to the energy dependence
of the hadron-nucleon parameters and propose a simple model which satisfies partial 
wave unitarity  and can be used to extrapolate these quantities to very high 
energy.
These  extrapolations are then used in section V to calculate the
hadron-air and nucleus-air cross sections in the energy range needed
in the phenomenology of extensive air showers produced by cosmic rays.  
In the conclusion we summarize our results and point out that more
hadron-nucleon data are needed to improve our understanding and aid
the analysis of cosmic ray showers. 

\section { Theoretical Formulation of Hadron-Nucleus  and Nucleus-Nucleus
Scattering}

Many authors including those listed in \cite{r1} have used the Glauber multiple 
diffraction approximation\cite{r2} to relate hadron-nucleus scatttering and 
nucleus-nucleus scattering to  the basic hadronic interaction.  In the 
interest of making this paper self contained we will review this procedure.

  The amplitude for hadron-nucleus elastic scattering can be written in 
the impact parameter representation as 
\begin{equation}
F_{h-A}(q^{2})={ik \over 2\pi}\int d^{2}\vec{ b} e^{i\vec{ q }\cdot 
\vec{ b}} (1-e^{i\chi_{h-A}})
\eqnum{2.1}
\end{equation}
where $\chi_{h-A }$ is the hadron-nuclear phase shift.  The 
quantity $\Gamma_{h-A}= (1-e^{i\chi_{h-A}})$ is the scattering amplitude in 
impact parameter space, often refered to as the nuclear profile function, 
and is bounded by unitarity because the imaginary part of  $\chi_{h-A }$ is 
positive.   In this treatment we will ignore spin and treat neutrons   
 and protons alike as nucleons.   
The essential approximation of the Glauber model is that the 
scattering is predominately forward and there is no momentum transfer to the individual nucloens.  In this limit the 
nuclear phase shift for a particular nucleon configuration is given by
the sum of the phase shifts for the collisions 
of the hadron with the individual nucleons in the nucleus.  We believe that this should 
give good results for the  total cross section given by the imaginary part of the forward amplitude and
for the elastic scattering which is dominated by the forward diffraction peak.  On the other hand, the fact that
the nuclear amplitude contains no nuclear recoil makes it unlikely that this approximation will give good results 
for quasi-elastic processes in which the nucleus ends up in an excited state or fragmented.  For cosmic ray applications
there are a number of processes that fall into the general catagory of target fragmentation that reduce the beam momentum, 
but do not contribute to shower formation.  Since these must be treated in the simulation, we see no particular reason to use the
Glauber model calculation of the quasi-elastic scattering to remove an ill definited and perhaps 
incorrect fraction of these events.

  If correlations 
between the nucleons are ignored and we average over the position of the 
(A) nucleons relative to the center of mass of the nucleus, the overall
hadron-nucleus phase shift is
\begin{equation}
e^{i\chi_{h-A}(\vec{b})}=\int\prod_{i=1}^A d^3r_i\rho
(\vec{r}_i)e^{i\sum_{j=1}^A\chi_{h-N}(\vec{b}-\vec{r}_{j\perp})}.
\eqnum{2.2}
\end{equation}
 The contribution 
of each nucleon to $\Gamma_{h-A}$ factorizes and we obtain
\begin{equation}
F_{h-A}(q^{2})={ik \over 2\pi}\int d^{2}\vec{ b} e^{i\vec{ q }\cdot 
\vec{ b}} \{1-[1-\int d^{3}r \Gamma_{h-N}(\vec{b}-\vec{r}_{\perp})
\rho(\vec{r})]^{A}\}
\eqnum{2.3}
\end{equation}
where $\rho(\vec{r})$ is the single nucleon density, $\vec{r}_{\perp}$ is the
component of $\vec{r}$ in the impact parameter plane and $\Gamma_{h-N}$ is the
hadron-nucleon profile function which is just the Bessel transform of the hadron-nucleon scattering amplitude:
\begin{equation}
\Gamma_{h-N}(\vec{ b})={1 \over 2\pi ik}\int d^{2}\vec{ q} e^{-i\vec{ q }\cdot 
\vec{ b}} F_{h-N}(\vec{q}).
\eqnum{2.4}
\end{equation}
All of the scattering processes are at a fixed incident hadron momentum.  
Being able to neglect the nuclear Fermi momentum is another simplification 
that is possible at high energy.

The multiple diffraction approximation can also be applied to nucleus-nucleus
scattering\cite{r7} (see also Refs. \cite{r5} and \cite{r4}).  In this case the
overall phase shift is, according 
to the Glauber assumption, simply the sum of the phase shifts of
$A\times A'$ 
individual nucleon-nucleon interactions.  In this case we must average over 
the positions of the nucleons in both nuclei. Unlike the
hadron-nucleus case, however, the integrals do not factorize and lead
to very complicated expressions even for the 2$\times$3 case. It can
be shown\cite{r7} that in the optical limit, that is both $A,A'\gg 1$, many
terms can be neglected and the elastic scattering amplitude can be  approximated by: 

\begin{equation}
F_{A'-A}(q^{2})={ik \over 2\pi}\int d^{2}\vec{ b} e^{i\vec{ q }\cdot 
\vec{ b}} \{1-[1-\int d^{3}r\int d^{3}r'\rho_{A}(\vec{r}) 
\Gamma_{N-N}(\vec{b}+\vec{r}_{\perp}-\vec{r}'_{\perp})
\rho_{A'}(\vec{r}')]^{AA'}\}.
\eqnum{2.5}
\end{equation}
In practice the optical limit is a good approximation for $A,A'\ge 10$\cite{r7}.
 From this expression, using the optical theorem, we otain the total cross section 
\begin{equation}
\sigma^{T}_{A-A'}={4\pi\over k}ImF_{A'-A}(0)
\eqnum{2.6}
\end{equation}
and the elastic cross section 
\begin{equation}
\sigma^{el}_{A-A'}=\int d\Omega_{k}\mid F_{A'-A}(q^{2})\mid^{2}.
\eqnum{2.7}
\end{equation}

In this work we will use a Gaussian single nucleon density  
\begin{equation}
\rho_{A}(\vec{r})={R(A)^{3}\over (2\pi)^{3\over2}} e^{-{r^{2}\over R(A)^{2}}}.
\eqnum{2.8}
\end{equation}
This is of course only valid for $A>1$ and for $A=1$ the density should be
a 
$\delta$-function.
While other
distributions such as Saxon-Wood are probably more realistic for large
$A$ and shell wavefunctions are more appropriate for small $A$, this form seems
 to work very well.  When the density given in Eq. 2.8 is  
combined with the observed form for the hadron-nucleon scattering amplitudes,
 Eqs. 2.6 and 2.7 can  be evaluated analytically.  We assume 
that $R(A)$ has the following form:
\begin{equation}
R(A)=R_{0}A^{\gamma}.
\eqnum{2.9}
\end{equation}
If we use experiment to determine $\Gamma_{h-N}$, the only free parameters in
this model are $R_{0}$ and $\gamma$. These could of course depend both on 
energy as well as particle type, although the simplest picture is one in 
which the 
density only reflects the average positions of the nucleons in the nucleus.

The hadron-nucleon elastic amplitude at small momentum transfer are well fit
by a Gaussian form.   More complicated dependence appears at larger 
angles, but it is  the simple small angle region that makes the dominant contribution to the
elastic cross sections.  This remains the case up to the highest experimental
energies available at the Tevatron collider.  The form of the 
scattering amplitude is then
\begin{equation}
F_{h-N}={k\sigma^{T}_{h-N}\over 4\pi}(\rho+i)e^{Bt\over2}
\eqnum{2.10}
\end{equation}
where B is the slope parameter $(t=-q^{2})$ and $\rho$ is the ratio of the 
real to imaginary part of the amplitude.  Here we have assumed that for 
small $q^{2}$,  $\rho$ is a constant.  At high energy $\rho$ is
quite small, typically 0.10-0.14, and we take it to be zero in our work.   The resulting hadron-nucleus absorptive cross section   
is 
\begin{equation}
\sigma^{abs}_{h-A}=\beta\int^{D}_{0}dx{{(1+x)^{2A}-1}\over x}=\beta\sum^{2A}
_{n=1}\left(\begin{array}{c} 2A\\n \end{array}\right){D^{n}\over n}
\eqnum{2.11}
\end{equation}
and the total cross section is 
\begin{equation}
\sigma^T_{h-A}=2\beta\int^{D}_{0}dx{{(1+x)^{A}-1}\over x}=2\beta\sum^{A}
_{n=1}\left(\begin{array}{c} A\\n \end{array}\right){D^{n}\over n}
\eqnum{2.12}
\end{equation}
where
\begin{equation}
\beta=-\pi(2B+R^{2}(A))
\eqnum{2.13}
\end{equation}
and
\begin{equation}
D={\sigma^{T}_{h-N}\over 2\beta}.
\eqnum{2.14}
\end{equation}
Finally, we include the prediction for the quasi-elastic scattering.  While we are doubtful about its reliability, it turns out be
be small and we will be force to use it in our fit to accelerator data.
\begin{equation}
\sigma^{qe}_{h-A}=\beta\int^{1}_{0}dx{{(1-Dx+{D^{2}(1+\gamma)^{2}\over(1+2\gamma)}x^{2+2\gamma \over (1+2\gamma)})^{A}-1}\over x} - \sigma^{abs}_{h-A}
\eqnum{2.15}
\end{equation}
where $\gamma={R^{2} \over 2B}$

The nucleus-nucleus cross sections can be obtained from Eqs. 2.11 and
2.12 by
letting $A\rightarrow AA'$ and using 
\begin{equation}
\beta=-\pi(2B+R(A)^{2}+R(A')^{2})
\eqnum{2.16}
\end{equation}
and
\begin{equation}
D={\sigma^{T}_{N-N}\over 2\beta}
\eqnum{2.17}
\end{equation}
The following identity\cite{r1} 
\begin{equation}
Lim_{M \rightarrow \infty
}  \int^{D}_{0}dx{{(1+x)^{M}-1}\over x}=Ei(MD)-\ln(-MD)-C
\eqnum{2.18}
\end{equation}
where $C $ is Euler's constant, can be used to evaluate these cross
sections for large $A$ or $AA'$.
Up to this point we have neglected position correlations of the
nucleons in the nucleus. Indeed, the nucleon coordinates $\vec{r}_i$ are
subject to the center of mass constraint ${1\over A}\sum_{i=1}^A \vec{r}_i=\vec{r}_
{cm}$. It
can be easily shown, by transforming to the cm coordinates
$\vec{r}_i^{cm}=\vec{r}_i-{1\over A}\sum_{j=1}^A \vec{r}_j$, that the effect of the cm
constraint on the elastic scattering amplitude, for the case of
Gaussian nucleon density distributions, is simply
\begin{equation}
F_{h-A}^{cm}(q)=e^{q^2{R(A)^2\over 4A}}F_{p-A}(q)
\eqnum{2.19}
\end{equation}
\begin{equation}
F_{A-A'}^{cm}(q)=e^{q^2({{R(A)^2}\over 4A}+{{R(A')^2}\over 4A'})}F_{A-A'}(q).
\eqnum{2.20}
\end{equation}
Thus the total cross sections are unaffected by the cm constraint. The
inelastic and elastic cross sections however are modified, although as
it is shown in Fig.1 the correction is at most a few mb and decreases as
$\sim A^{-1/3}$. We find
\begin{equation}
\Delta\sigma^{inel}_{p-A}=-{2\beta\over A}\sum_{N=1}^{A}\sum_{M=1}^A\left(
\begin{array}{c}A\\ N\\ \end{array}\right)
\left( \begin{array}{c}A\\M\\ \end{array}\right) D^{M+N}
{{MN}\over{(M+N)[(M+N)({{2B+R(A)^2}\over {R(A)^2}})-{2MN\over A}]}}
\eqnum{2.21}
\end{equation}
and for $A\gg 1$
\begin{equation}
\Delta\sigma^{inel}_{p-A}\simeq {1\over 2A}[\sigma^{inel}_{p-A}+\beta
(1+D)^{2A}].
\eqnum{2.22}
\end{equation}

 A recent model developed by Engel et al. (hereafter the EGLS model) 
\cite{r5} and discussed in more detail
 by Fletcher et al.\cite{fletcher} is based in part on the work of Durand 
and Pi 
\cite{r6}, which is a semi-classical treatment motivated by the parton model.  
The EGLS model is a
probabilistic approach to hadron-nucleus and nucleus-nucleus inelastic
scattering, treating the individual hadron-nucleon interactions
incoherently, and summing probabilities of interaction instead of
amplitudes.  While this treatment is somewhat different from ours, for  
the special case of purely imaginary nuclear 
phase shift the results are identical.

\section{cross section fits}

The input parameters of Glauber's model are the hadron-nucleon cross
sections $\sigma^T_{h-N}$ and nuclear slopes $B_{h-N}$. We determine
these at energies of 60, 200, and 280 GeV. The total cross sections are
found in a compilation of data\cite{r12}. 
Some sample fits are, ($s$ is in GeV$^2$ and $\sigma^T$ is in mb)  
\begin{equation}
\sigma^T_{\pi^+-p}=23.1+.525(\ln{s/112})^2
\eqnum{3.1}
\end{equation}
and
\begin{equation}
\sigma^T_{\pi^--p}=24.0+.582(\ln{s/170})^2
\eqnum{3.2}
\end{equation}
with a $\chi^2={1.12\over d.f.}$ and $\chi^2={1.30\over d.f.}$
correspondingly. There are some values of the slope parameters in the
literature \cite{r8,r9,r10} but we obtain most values by fitting directly the elastic
differential cross section data (at low momentum transfer) with an
exponential form:
\begin{equation}
{{d\sigma^{el}_{h-N}}\over{dt}}=Ae^{Bt}.
\eqnum{3.3}
\end{equation}
Sample fits of these values are, ($s$ is in GeV$^2$ and $B$ is in GeV$^{-2}$)   
\begin{equation}
B_{\pi^--p}=9.53-0.705\ln s+0.136(\ln s)^2
\eqnum{3.4}
\end{equation}   
\begin{equation}
B_{\pi^+-p}=7.90-0.155\ln s+0.076(\ln s)^2.
\eqnum{3.5}
\end{equation}   
Ayres et al. \cite{r11} have measured $\pi-p$ and $K-p$  differential 
elastic cross sections  at
50, 70, 100, 140, 175, and 200 GeV.  There are some 
data for $\pi^--p$ at 360 GeV \cite{r16} and for $\pi-p$ at 200 GeV \cite{r15}. By interpolating and
extrapolating in the case of $K-p$ we obtain $B_{h-p}$ at the required
energies. The only free parameters in our model are $R_0$ and $\gamma$
that appear in the nucleon density distribution.  The most recent hadron-nucleus data is due to
 Carroll et al.\cite{r3}.  The comparison of our model with their data is complicated by the fact that they removed the 
quasi-elastic cross section from their absorption cross section.  While this is a small correction to the absorption data it was done
in the process of extrapolating the data to zero momentum transfer and  hence we cannot obtain the total absorption cross sections from 
their published data.  As a result, we have two courses of comparison.  First, we can simply ignore these small corrections and fit their
data with our model.  The alternative is to use the Glauber model to calculated the quasi-elastic cross sections.  
This has the disadvantage of subtracting a small but unreliable term from the quantities that we believe can be calculated accurately.   

 We first ignore the quasi-elastic cross sections and simply  fit the  108 data points of  Carroll et al.\cite{r3} for scattering of $p,
\bar{p}, K^+, K^-, \pi^+, \pi^-$ off Li, C, Al, Cu, Sn, and Pb targets
at 60, 200, and 280 GeV incident hadron energy with our two free parameters.
These results are shown in Figs. 2 and 3.  We obtain,
\begin {equation}
R_0=3.89\hspace{0.1in} GeV^{-1}
\eqnum{3.6}
\end{equation}
\begin {equation}
\gamma=0.31
\eqnum{3.7}
\end{equation}
with $\chi^2/d.f.=0.25$. This  $\chi^2$ was calculated using the quoted error 
of the data which includes an estimated systematic error of 3\%.  The small 
value of our $\chi^2$ is an indication that our fit is not sensitive to the 
type of systematic error that exists in the data.  
 If we use only the statistical error
of 1\% as  quoted by Carroll et al.\cite{r3} to calculate $\chi^2$ we obtain a 
$\chi^2$
$\sim
2$ per degree of freedom.

If we use the Glauber model to remove the quasi-elastic scattering from our fit, we obtain,
\begin {equation}
R_0=4.74\hspace{0.1in} GeV^{-1}
\eqnum{3.8}
\end{equation}
\begin {equation}
\gamma=0.28
\eqnum{3.9}
\end{equation}
with $\chi^2/d.f.=0.34$.  While this is not as good a fit as the one described above, it is quite acceptable.
Note that there is a noticeable change in the parameters.

\section { Unitarity constraint at high energy}

The data for the total nucleon-nucleon cross section $\sigma^T$ and
nuclear slope $B$ up to Tevatron energies \cite{r15} can be fit with quadratic
logarithmic forms and at infinite energy they grow as $\ln(s)^2$.
Our fits, shown in Figs. 4 and 5  are: ($s$ is in GeV$^2$, $\sigma^T$ in mb and $B$ in GeV$^{-2}$)
\begin{equation}
\sigma^T_{p-p}=38.46+0.41(\ln(s/118))^2
\eqnum{4.1}
\end{equation}
and    
\begin{equation}
B_{p-p}=9.25+0.37\ln(s)+0.01(\ln(s))^2
\eqnum{4.2}
\end{equation}
where we have assumed that at $p-p$ and $p-\bar{p}$ are identical at collider energies. 
Note that for $s\rightarrow\infty$, $\Gamma(0)={\sigma^T\over4\pi B}\rightarrow
8.35$. In fact, the unitarity limit is saturated at
$\sqrt s\simeq2.5$ TeV (see Fig. 6). To prevent unitarity violation
we suggest that at high energies the elastic scattering amplitude
deviates from the pure exponential form in $t$ that is commonly used
at small scattering angles. We set,
\begin{equation} 
f(t,s)={ik\over4\pi}\sigma^T(s)e^{B(s)t\over 2}J_0[\alpha(s)\sqrt{-t}]
\eqnum{4.3}
\end{equation}
Since ${dJ_0(x)/dx}|_{x=0}=0$, $B$ is still defined as the
nuclear slope of the elastic amplitude. The profile function that
follows from Eq. 4.3  becomes smeared with a modified Bessel function:
(see Fig. 6 )
\begin{equation}
\Gamma(b)={\sigma^T\over{4\pi
B}}e^{{\alpha^2+b^2}\over2B}I_0({\alpha b\over B})
\eqnum{4.4}
\end{equation}
and $\alpha(s)$ is chosen such that $\Gamma(b,s)\leq1$ or
$\alpha\geq[2B\ln (\frac{\sigma^T}{4\pi B})]^{1/2}$. For $\sqrt s\leq
2.5$ TeV we set $\alpha =0$ so that our cross section fits are not
modified since they are well below this energy. At higher energies the
above profile function leads to  
\begin{equation}
\sigma^T_{h-A}=-2\beta \int_0^\infty dx\{1-[1+D'e^{-x}I_0(\lambda\sqrt x)]^A\}
\eqnum{4.5}
\end{equation}
\begin{equation}
\sigma^{inel}_{h-A}=-\beta \int_0^\infty dx\{1-[1+D'e^{-x}I_0(\lambda\sqrt x)]^{2A}\}
\eqnum{4.6}
\end{equation}
where $\beta=-\pi(2B+R(A)^2)$, $\lambda=2\alpha\sqrt{\pi/-\beta}$ and
$D'=De^{\lambda^2/4}$.
The nucleus-nucleus case can be obtained as before, in the optical limit, by
substituting $A\rightarrow AA'$ and $\beta=-\pi(2B+R(A)^2+R(A')^2)$. As
it is shown in Fig. 6  for $A=Air$, the $p-A$ profile function remains
unchanged for small impact parameter, (saturating the unitarity limit
for large $A$) but increases at large $b$, due to the smearing of the
nucleon-nucleon profile function. This results in an increase of the total and
inelastic $p-A$ cross sections, as shown in Fig. 7.    In these plots we have
used the minimum value of $\alpha (s)$ that satisfies unitarity.

Our model of course is by no means unique since there may be other ways to
preserve unitarity.
The choice of the Bessel function is not essential and other
modifications such as exponentials etc. should produce the same
qualitative results.

\section{Extrapolation to cosmic ray energies}

The development of cosmic ray showers in the atmosphere depends
strongly on the size of the relevant hadron-nucleus and
nucleus-nucleus cross sections at very high energies $E\sim 10^{18}$
eV, $\sqrt{s} \sim 40$ TeV. Both the interaction length in the
atmosphere and the number of nucleons in the nucleus that participate
in a collision are a function of the absorption cross sections.

In the previous sections we have shown that the hadron-nucleus cross
sections can be calculated reliably from the basic hadron-nucleon
interaction. Since Glauber's analysis does not depend explicitly on
the energy scale at which the scattering takes place we expect that it
is still applicable at very high energy.  On the other hand, only for $p-\bar{p}
$ does experimental data approach modest cosmic ray energies.  Even this data 
at $\sqrt{s}=1.8$ TeV ($E_{lab}\simeq 1.7 \times 
10^6$ GeV)  must be extrapolated over three decades in energy to reach the 
highest cosmic ray energies of  $E_p>10^9$ GeV.  Because of the observed 
logarithmic behavior of the cross section and slope parameter, this procedure
should be fairly reliable and data that will eventually come from the LHC 
should provide direct input for all but the very highest energies.  The situation 
for the  $K^{\pm}-N$ and $\pi^{\pm}-N$ interactions is much less satisfactory.  
Data for these reactions exist only up to a few hundred GeV lab energy with 
little hope of future measurements above the TeV range even with the LHC.  
While external beam experiments up to the Tevatron's maximum energy would 
improve
our understanding of these reactions, it
is unlikely that energy extrapolations over 7 or 8 orders of 
magnitude can be trusted, which means that the forms given in Eqs. 3.1 
through 3.4 cannot be used at cosmic ray energies.  It appears that the best
method for determining  $K^{\pm}-N$ and $\pi^{\pm}-N$
scattering at cosmic-ray energies is to use the independent quark model result  

\begin{equation} 
\sigma^T_{\pi -N}=\sigma^T_{K-N}={2\over 3}\sigma^T_{N-N}
\eqnum{5.1}
\end{equation}
Although this is only approximately true at low energy, it is expected
to improve at very high energy. The hadronic nuclear slopes are taken
equal. These approximations are probably adequate considering the fact that the hadron-nucleus
cross-sections are relatively insensitive to the precise value of $B_{h-N}$ and
$\sigma^T_{h-N}$ as it is shown in Figs. 8 and 9  for  $p-Air$.
In Fig. 7 we plot our prediction for the $p-Air$ cross section as a
function of center of mass energy and compare with a fit by
Block and Cahn of \cite{r14}.   Based on theoretical arguments they assumed
that the total proton-proton cross section has the form
\begin{equation}
\sigma^T_{p-p}=A+\beta\left(\ln^2({s\over s_0})-{\pi^2\over
4}\right)+c\sin({\pi\mu\over 2})s^{\mu -1}+D\cos({\pi\alpha\over
2})s^{\alpha -1}
\eqnum{5.2}
\end{equation}
The parameters from their fit are $A=41.74$ mb, $\beta =0.66$ mb, 
$s_0=338.5$ GeV$^2$, $c=0$,
$D=-39.37$ mbGeV$^{2(1-\alpha)}$, and $\alpha =0.48$.
The nucleus-nucleus cross sections can also be calculated at cosmic
ray energies based on the nucleon-nucleon interaction using Eqs. 2.11
and 2.12 with the appropriate substitutions. However, there are not enough
experimental data on nucleus-nucleus cross sections  to check whether
the Glauber model works as well as in the hadron-nucleus case,
especially since for practical purposes one is forced to use the
optical limit approximation.

\section{conclusion} 

We have shown that Glauber's model of multiple diffraction accurately predicts
hadron-nucleus cross sections from the basic underlying hadron-nucleon
interaction. We have obtained excellent agreement with the
experimental data for a variety of incident hadrons and target nuclei
in the energy range of 60-280 GeV with just two free parameters.
External beam experiments at 800 GeV on nuclear targets would provide a 
useful check of the validity of these methods over a much broader range of 
energies.
Since Glauber's analysis does
not depend explicitly on the energy scale of the scattering process and it is
unlikely that the distribution of nucleons inside a nucleus would change 
with energy, we expected  this method to continue to be applicable at very 
high energies. The $p-\bar p$ data extend to
a large energy range and  can be extrapolated to cosmic ray energies.
Using Glauber's model we
can then obtain the relevant nucleon-nucleus cross sections at these
energies. We can use the independent quark model to  estimate the  $\pi -N$ and $K-N$ cross sections, but have not yet developed a model for the nuclear 
slope. 

Finally,  we would like to emphasize that extrapolating our fits to
available data on $\sigma^T_{p-p}$ and $B_{p-p}$ to energy $\sqrt{s}$ greater than
a few TeV violates unitarity.  This means that above the current Tevatron 
energies and the energy available at the LHC we should expect some change
in the shape of the nucleon diffraction peak and its related profile function. 
 This is to be contrasted with the behavior at lower energies where the 
Gaussian profile function height and radius have grown logarithmically with s.

We propose a simple model in which the
p-p elastic scattering amplitude deviates from a pure exponential in the
momentum transfer and satisfies unitarity. The resulting absorption
cross sections are increased appreciably at cosmic ray energies.  How nature
actually chooses to satisfy unitarity awaits experiments in the few TeV energy range.

\begin{figure}
\caption{Correction to the $p-A$ inelastic cross-section due to the
center of mass constraint as a function of $A$ at $\protect\sqrt{s}=10^2$GeV. 
The dotted line is the large $A$ approximation.
}
\end{figure}
\begin{figure}
\caption{Fits to $p-A$, $\pi^+-A$, and $K^+-A$ absorption cross sections
at incident hadron momenta of 60, 200, and 280 GeV/c. Shown are the
data by Carroll et al. and our fitted values connected by straight
line segments.} 
\end{figure}
\begin{figure}

\caption{Fits to $\bar{p}-A$, $\pi^--A$, and $K^--A$ absorption cross sections
at incident hadron momenta of 60, 200, and 280 GeV/c. Shown are the
data by Carroll et al. and our fitted values connected by straight
line segments.} 
\end{figure}
\begin{figure}

\caption{The $p-p$ total cross section as a function of $s$. The solid
line is our fit of Eq.4.1, and included  are the high energy $p-\bar{p}$
collider data which we assume are equal to $p-p$ at these energies.}
\end{figure}
\begin{figure}

\caption{The $p-p$ nuclear slope $B$ as a function of $s$. The solid
line is our fit of Eq.4.2, and included  are the high energy $p-\bar{p}$
collider data which we assume are equal to $p-p$ at these energies.}
\end{figure}
\begin{figure}

\caption{Profile  functions for $p-p$ and $p-Air$ scattering at $s=10^9$ GeV$^2$. 
Shown are $\Gamma_{p-p}$ the unitarity violating profile, the
Bessel-modified profile $\Gamma^B_{p-p}$ and the $p-Air$ profiles
that follow from them using Glauber's approximation.}  
\end{figure}
\begin{figure}

\caption{Inelastic $p-Air$ cross sections as a function of cm energy. The
solid line is the Bessel-modified cross section which is higher than
the unitarity violating cross section above $\sim 2.5$ TeV (dashed segment). Also
shown for comparison as a dashed line, is the Block and Cahn parametrization of Eq. 5.2}
\end{figure}

\begin{figure}
\caption{Dependence of the $p-Air$ inelastic cross section on the nuclear
slope of the proton-proton elastic scattering, keeping the $p-p$ cross
section constant at 40 mb.}
\end{figure}
\begin{figure}

\caption{Dependence of the $p-Air$ inelastic cross section on the proton-proton
cross section, keeping the $p-p$ nuclear slope constant at 10 and 20 $GeV^{-2}$.}
\end{figure}


\begin{references}
\bibitem{r1}Victor Franco, Phys. Rev.~ Letters {\bf 24},1452 (1970);
Victor Franco, Phys. Rev.~ Letters {\bf 32}, 911 (1974); Paul M. Fishbane and J.
S. Trefil, Phys. Rev.~ Letters  {\bf 32},
396 (1974)

\bibitem{r2} R.J. Glauber, in{ \sl Lectures in Theoretical Physics}, edited by
 W. Britten and L. G. Dunham (Interscience, New York,1959)Vol. 1, p.253; R.J.
 Glauber and G.Matthiae, Nucl. Phys. {\bf B21} (1970) 135.

\bibitem{r3} A. S. Carroll et al.,Phys. Letters {\bf 80B},319 (1979).
Denisov et al. of Ref. \cite{r9}  have also collected  hadron-nucleus data but
they are not in agreement with Carroll's et al.



\bibitem{r5} J. Engel et al., Phys. Rev. {\bf D46}, 5013 (1992);

\bibitem{fletcher} R. S. Fletcher et al.,  Phys. Rev. {\bf D50}, 5710 (1994).

\bibitem{r6} Loyal Durand and Hong Pi, Phys. Rev. {\bf D38}, 78 (1988).

\bibitem{r4} A. Bialas et al., Nucl. Physics {\bf B111}, 461 (1976), K. Kinoshita, A.
Minaka, and H. Sumiyoshi, Z. Phys. {\bf C8}, 205 (1981) and references
therein.  S. Date and D. Kiang, Phys. Rev. {\bf D36}, 2744 (1987).

\bibitem{r7} W. Czyz and L. C. Maximon, Ann. of Phys. {\bf 52}, 59
(1969).

\bibitem{r12} {\sl Total Cross Sections for Reactions of High Energy
Particles.} Landolt B\"{o}rnstein, New Series {\bf I/12a} and {\bf I/12b}, edited
by H. Schopper (1988).

\bibitem{r8} N. Amos et al., Nucl. Phys. {\bf B262}, 689 (1985).

\bibitem{r9} S. P. Denisov et al., Nucl. Phys. {\bf B61}, 62 (1973).

\bibitem{r10} M. Bozzo et al. (UA4 collaboration), Phys. Letters {\bf
B147}, 385 (1984); N. Amos et al., Phys. Rev. Letters {\bf 68}, 2433
(1992); C. Augier et al. (UA4/2 collaboration), Phys. Letters {\bf
B315}, 503 (1993); Phys. Letters {\bf B316}, 448 (1993).



\bibitem{r11} D. S. Ayres et al., Phys. Rev. {\bf D15}, 3105 (1977).

\bibitem{r16} A. Firestone et al.,Phys. Rev. {\bf D14}, 2902 (1976).

\bibitem{r15} A. Schiz et al.,Phys. Rev. {\bf D24} 26, (1981).


\bibitem{r13} N. A. Amos et al., Phys. Rev. Letters {\bf 63}, 2784 (1989);
N. A. Amos et al. Phys. Rev. Letters {\bf 61}, 525 (1988).  

\bibitem{r14} M. M. Block and R. N. Cahn, Rev. Mod. Phys. {\bf 57}, 563
(1985).


\end{references}
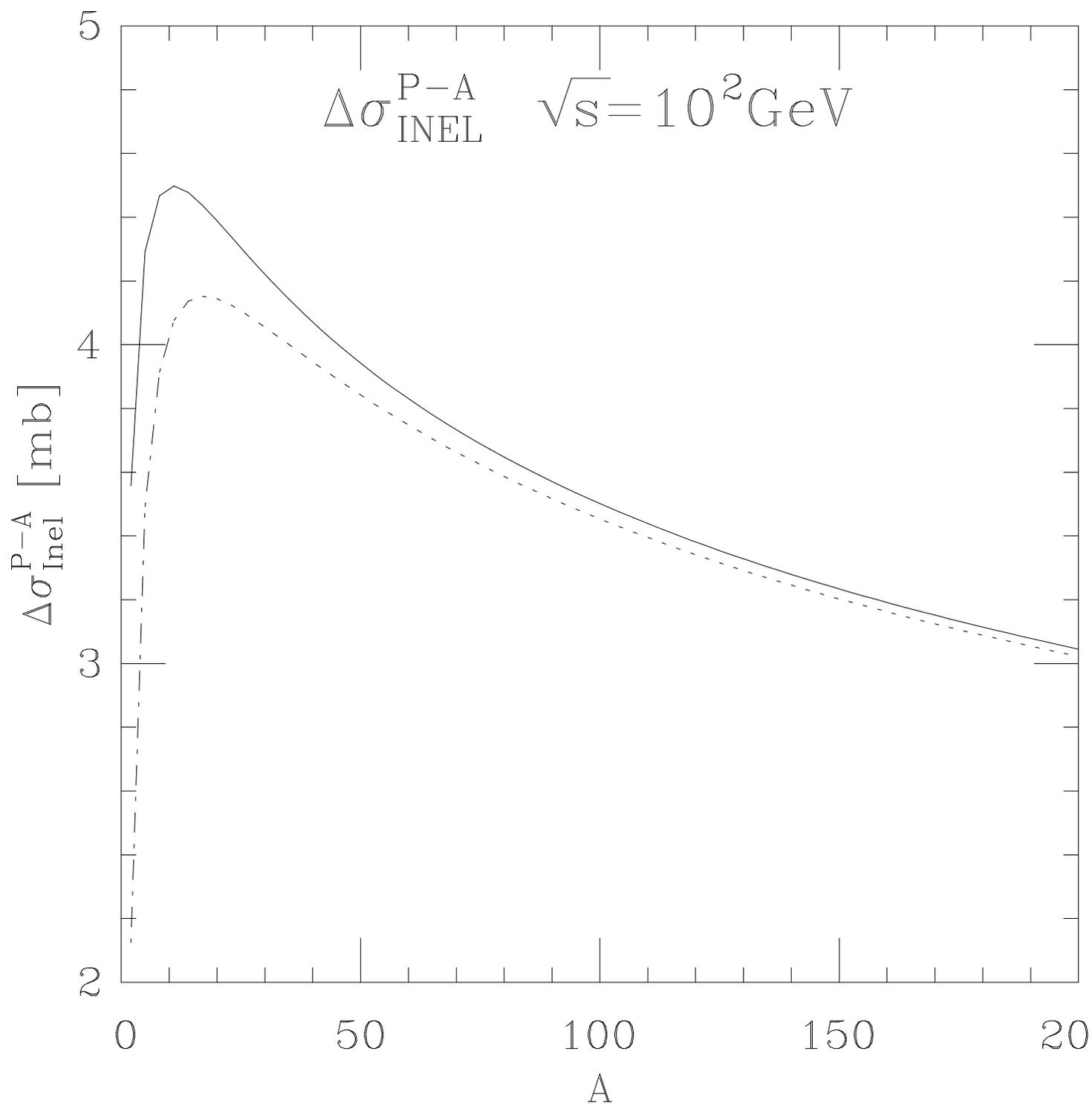
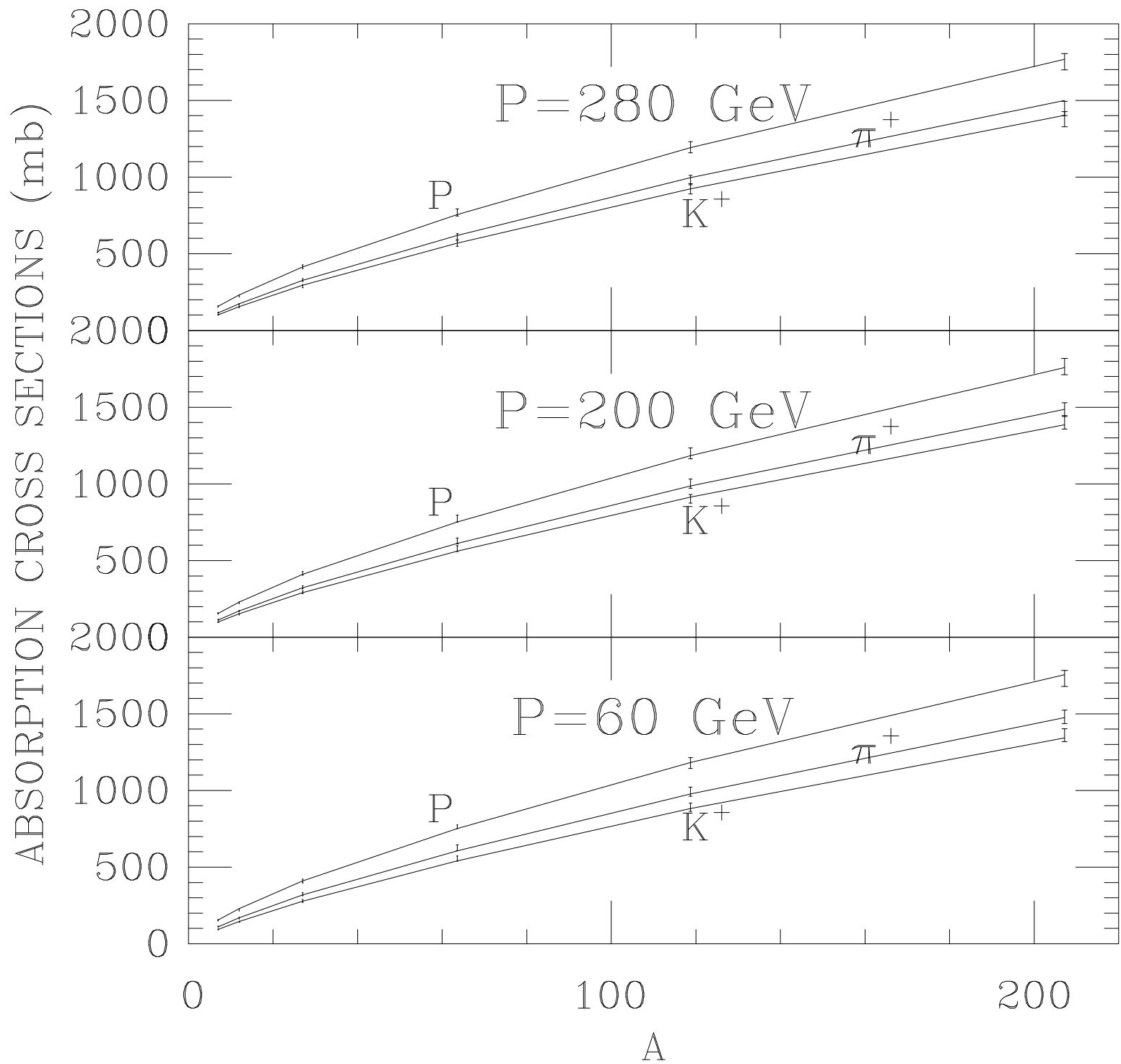
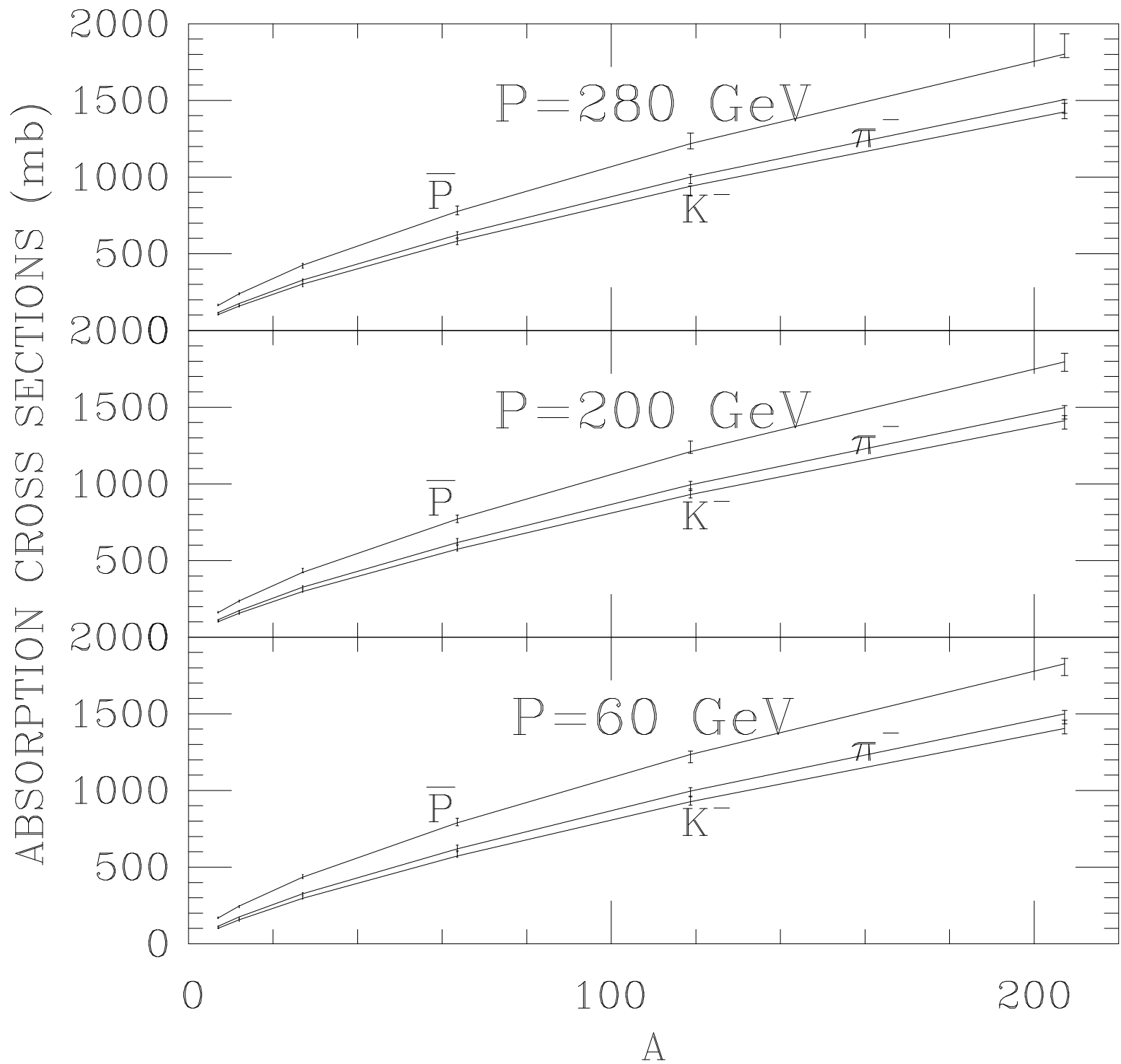
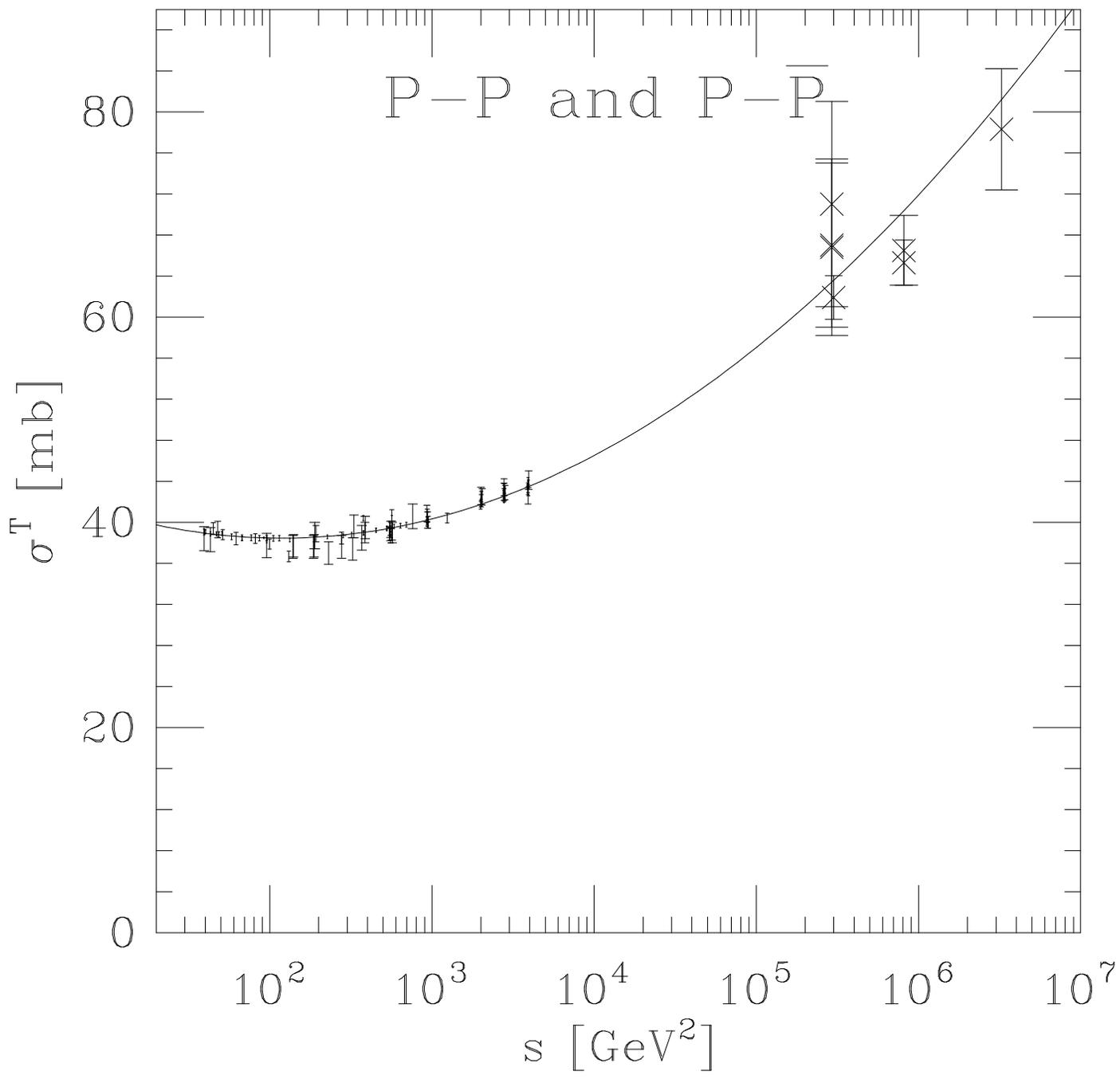
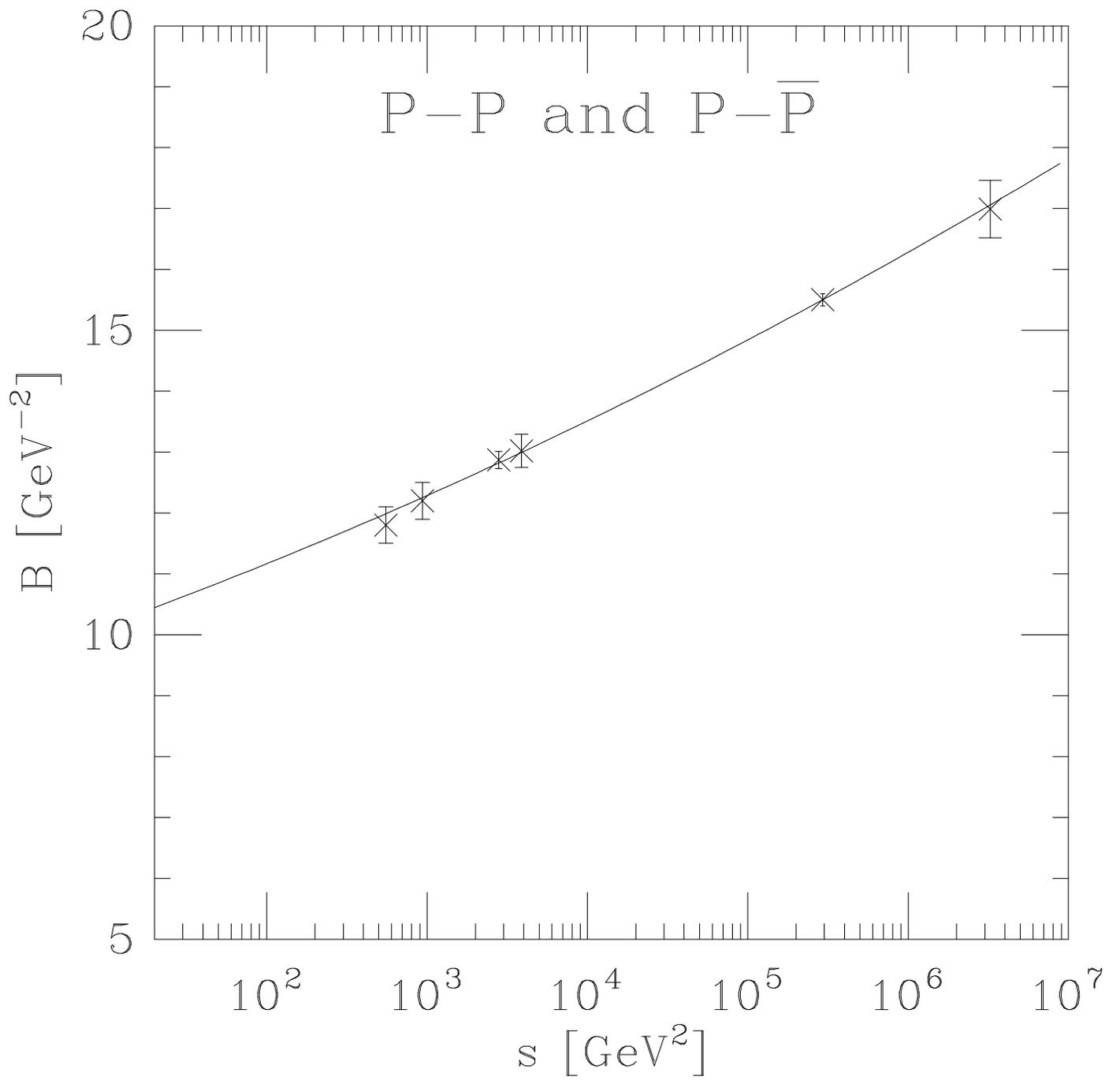
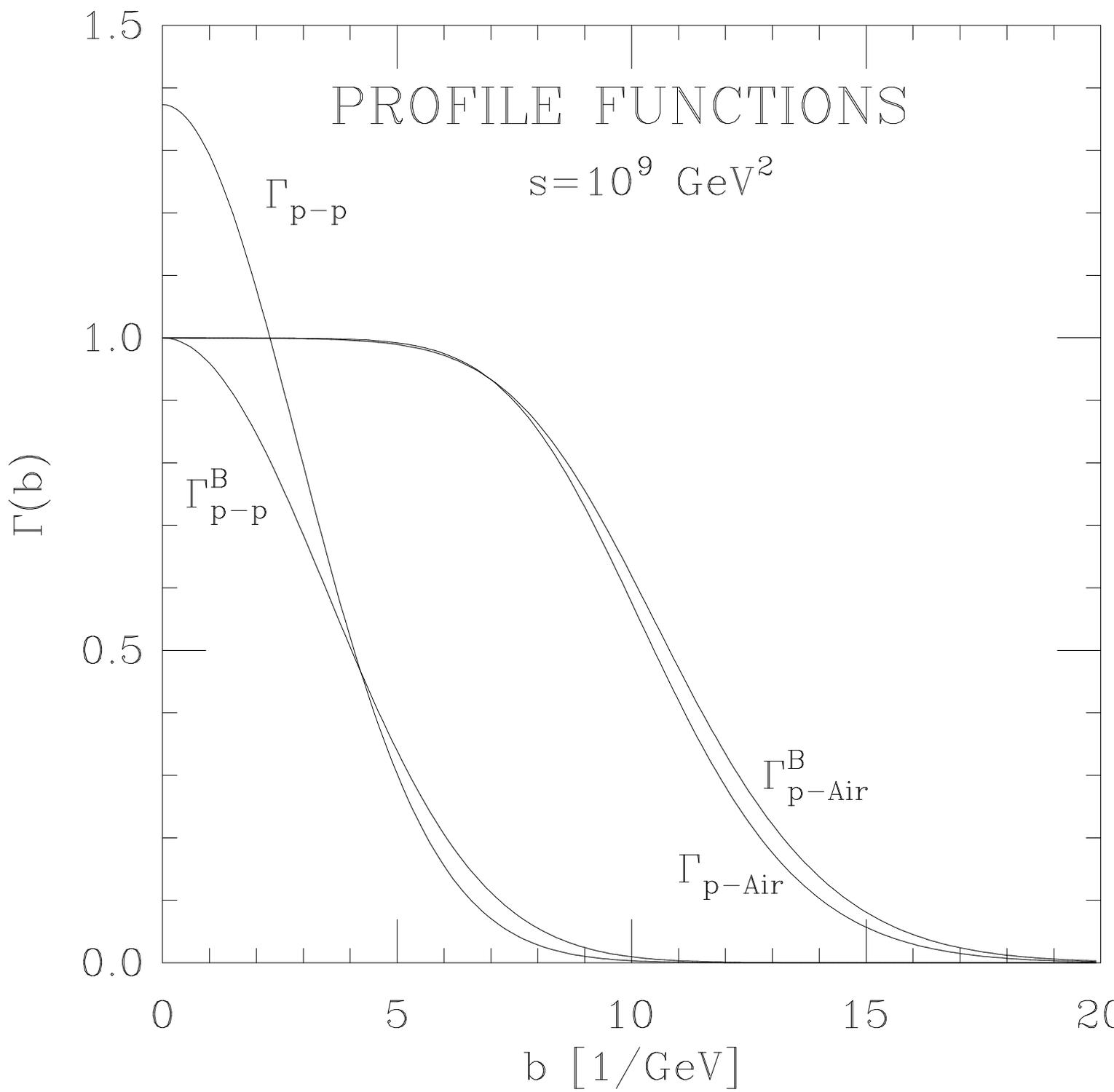
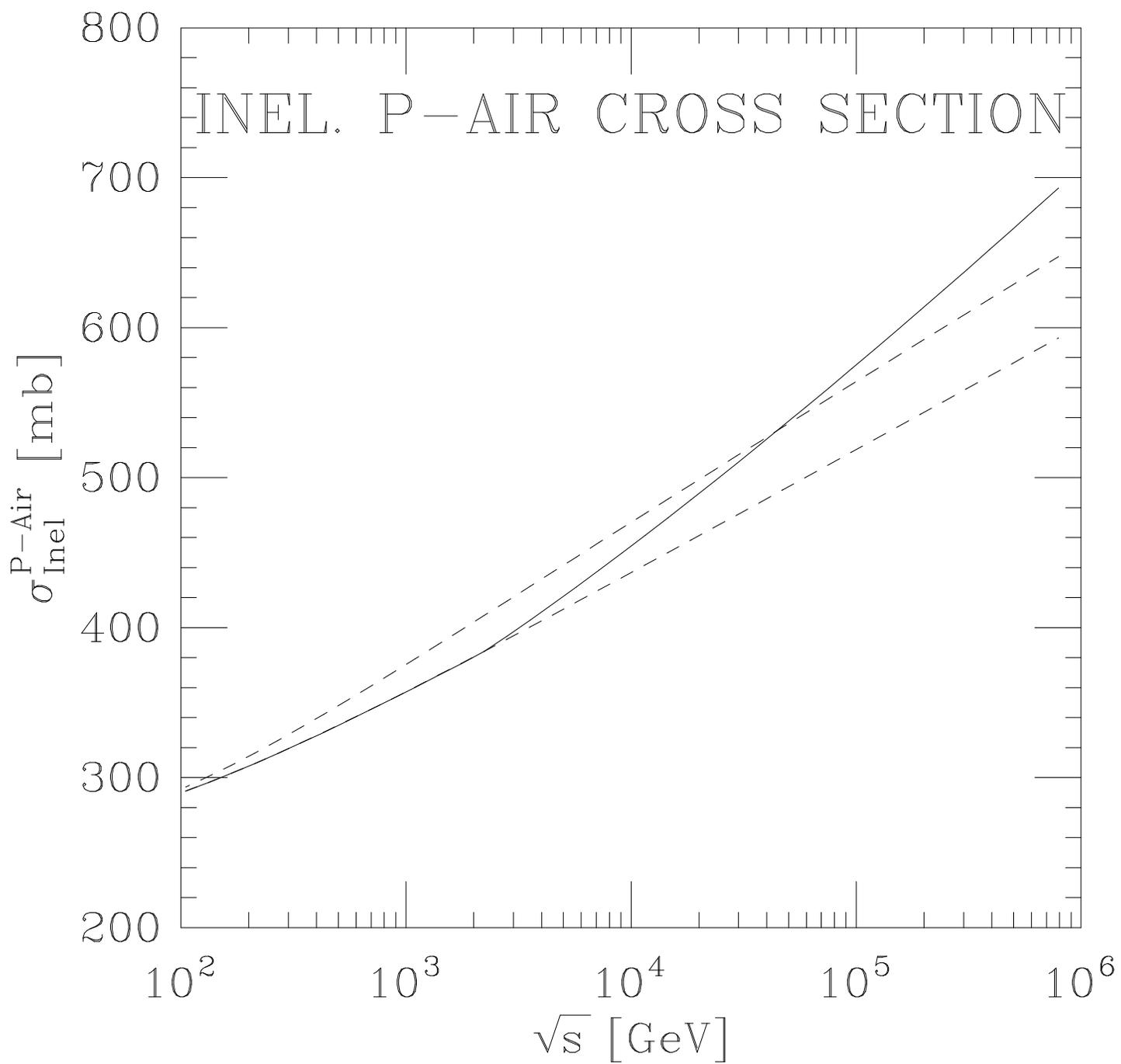
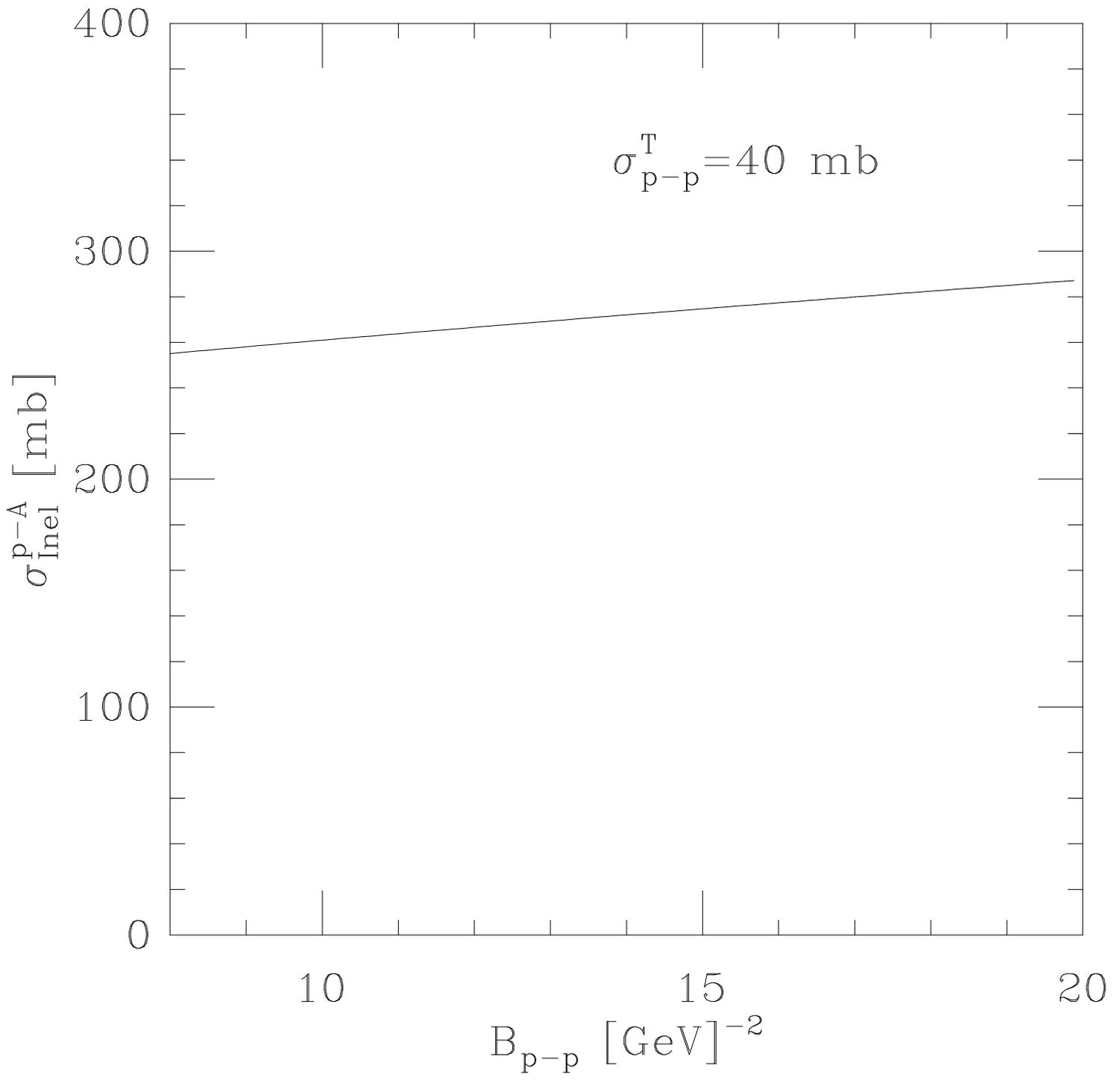
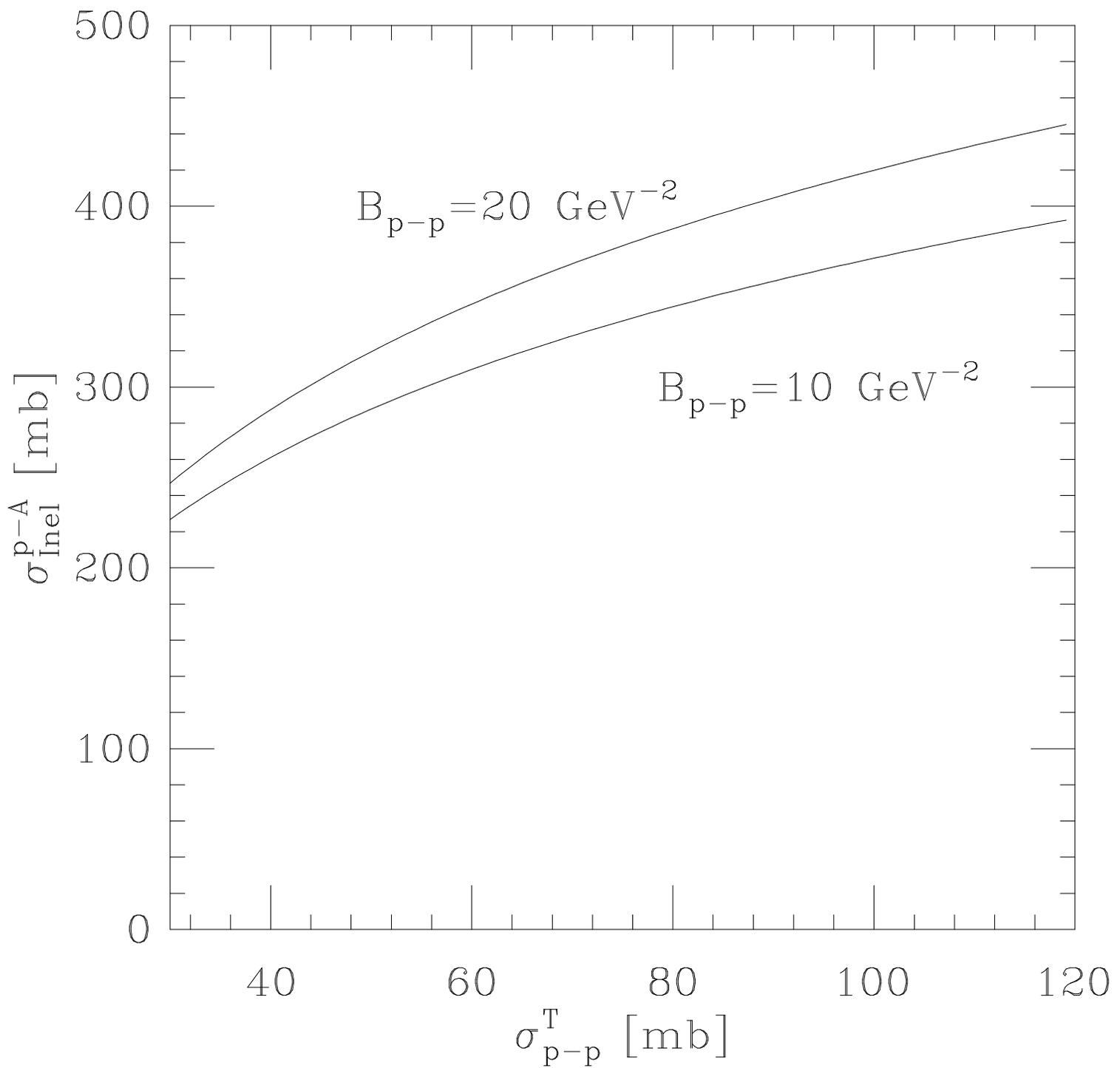
\end{document}